\documentclass[%
reprint,
superscriptaddress,
 amsmath,amssymb,
 aps,
]{revtex4-2}
\usepackage{graphicx}
\usepackage{dcolumn}
\usepackage{bm}
\usepackage{hyperref}
\usepackage[mathlines]{lineno}
\usepackage{natbib}
\usepackage{jabbrv}
\usepackage{multirow}
\usepackage{tabularx}

\begin{document}

\title{Ultracompact Low-Loss Grating Couplers}

\author{Shiang-Yu Huang}
\email{Corresponding author: shiang-yu.huang@fmq.uni-stuttgart.de}
\affiliation{Institute for Functional Matter and Quantum Technologies, University of Stuttgart, 70569 Stuttgart, Germany} 

\author{Jonas Zatsch}
\affiliation{Institute for Functional Matter and Quantum Technologies, University of Stuttgart, 70569 Stuttgart, Germany} 

\author{Tim Engling}
\affiliation{Institute for Functional Matter and Quantum Technologies, University of Stuttgart, 70569 Stuttgart, Germany} 

\author{Jeldrik Huster}
\affiliation{Institute for Functional Matter and Quantum Technologies, University of Stuttgart, 70569 Stuttgart, Germany} 

\author{Stefanie Barz}
\affiliation{Institute for Functional Matter and Quantum Technologies, University of Stuttgart, 70569 Stuttgart, Germany}
\affiliation{Center for Integrated Quantum Science and Technology (IQST), University of Stuttgart, 70569 Stuttgart, Germany}


\begin{abstract}
Fiber-to-chip couplers play a crucial role in interfacing on-chip photonic circuits with other optical systems or off-chip devices.
Downsizing the couplers via topology optimization addresses the demand for high-density integration and improves the scalability of photonic integrated systems. 
However, these optimized couplers have yet to reach the performance level demonstrated by their conventional counterparts, leaving room for further improvement.
In this work, we apply topology optimization to design single-polarization 1D and dual-polarization 2D grating couplers incorporating bottom reflectors and achieve sub-decibel coupling efficiency.
Both types of couplers are fabricated on the silicon-on-insulator platform with dimensions of merely 14~\textmu m~\texttimes~14~\textmu m and are compatible with standard single-mode fibers at normal incidence.
From our experimental characterization, the measured peak coupling efficiency of the topology-optimized 1D and 2D couplers is $-0.92(1)$ dB and $-0.86(13)$ dB, respectively, within the telecom C-band.
Our demonstration provides a coupling solution for photonic applications requiring high efficiency and high-density integration, such as spatial division multiplexing and photonic quantum technologies.
\end{abstract}

\maketitle

\begin{figure*}[t!]
  \centering
  \includegraphics[scale=0.75]{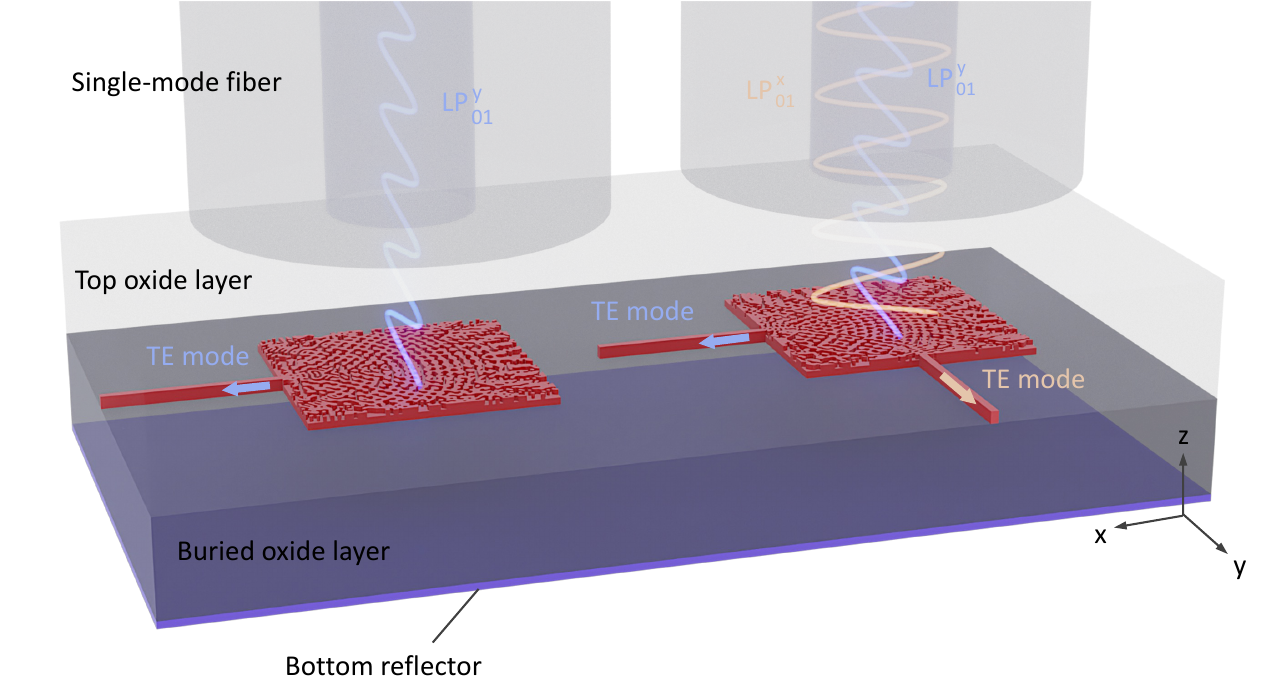}
  \caption{
    Schematic of the designed ultracompact topology-optimized 1D (left) and 2D grating couplers (right) incorporating a bottom reflector on the SOI platform.
    The topology-optimized 1D grating coupler couples only one specific linear polarization state (LP$^{y}_{01}$).
    In contrast, the topology-optimized 2D grating coupler couples light with arbitrary polarization (i.e. the superposition state of LP$^{x}_{01}$ and LP$^{y}_{01}$) into two waveguides with a proportion determined by the polarization state.
  }
  \label{fig:TOC}
  \end{figure*}

\section*{Introduction}
Photonic integrated systems have emerged as an advantageous platform for realizing various optical applications owing to their compactness, stability and scalability.
Such systems generally rely on fiber-to-chip couplers, for example edge \cite{hatori2014hybrid,he2020low,mu2020edge} and grating couplers \cite{zaoui2014bridging,hoppe2020ultra,liu2022high}, for establishing links between them and other on-chip systems or off-chip devices via optical fibers.
Edge couplers usually have higher coupling efficiency with broader operating bandwidth compared to grating couplers.
In contrast, grating couplers can be flexibly positioned in the chip layout due to their out-of-plane coupling scheme.
Therefore, they are suitable for wafer-scale testing as they provide access at nearly every location on a wafer.
Furthermore, grating couplers operate without special treatments of the chip facets that are required for edge couplers, greatly reducing the complexity of the fabrication process.

A variety of grating couplers have been demonstrated for a broad range of photonic applications.
The most basic and widely used one is the single-polarization 1D grating couplers \cite{zaoui2014bridging,hoppe2020ultra,liu2022high}, hereafter referred to as the 1D grating couplers.
The grating structure extending along solely one direction alters the refractive index to fulfill the Bragg condition and diffracts light towards the silicon waveguide.
To mitigate the mode-mismatch loss between the wide waveguide that hosts the grating structure and the narrow waveguide connecting to other circuit components,
one may introduce a long linear adiabatic taper or design the gratings in a focusing configuration \cite{hoppe2020ultra}.
Still, since the 1D grating structure cannot effectively diffract light with different polarization states at the same time, such couplers only achieve efficient coupling for one specific linear polarization (LP).
Hence, when coupling light with arbitrary polarization is necessary, dual-polarization 2D grating couplers, hereafter referred to as the 2D grating couplers, become a common solution \cite{luo2018low,watanabe20192,zhou2025efficient,luo2025silicon}. 
Conceptually, a 2D grating coupler can be considered as a device formed by superimposing two 1D grating couplers of which the connecting waveguides extend along two orthogonal directions. 
Light coupled by the 2D grating coupler is then guided into these two waveguides in a proportion depending on its polarization, which is essentially a transformation from the polarization modes into the spatial modes.

In recent years, using topology optimization to design polarization insensitive out-of-plane couplers \cite{kuster2025three}, 1D \cite{hammond2022multi,wang2024single,pita2025inverse} and 2D grating couplers \cite{hammond2022multi,wu2025ultra} has attained growing interest as it provides a promising route toward further miniaturization of photonic integrated components.
In topology optimization, the materials within a discretized design region are represented by the parametrized pixels and the optimization algorithm searches for a final structure that maximizes the device performance.
In this case, one can design grating couplers of which the main body straightforwardly connects to the integrated waveguides to have a minimal spatial footprint.
Without the long tapers, such a size reduction is about one order of magnitude.
Even when compared with a focusing grating coupler, the footprint is still reduced by roughly a factor of four.
However, despite the ongoing development, none of the abovementioned topology-optimized couplers have yet reached the sub-decibel coupling efficiency that state-of-the-art conventional grating couplers achieve (Table.~\ref{table: 1D coupler comparison} and~\ref{table: 2D coupler comparison}).
This may be attributed to the relatively strong transmission traveling through the substrate \cite{hammond2022multi,pita2025inverse} and the strong back reflection owing to the single fully-etched grating structures \cite{wang2024single, wu2025ultra}.
Such a level of inefficiency is particularly unfavorable in applications, for example, photonic quantum technologies, where high photon loss poses a major challenge.

In this work, we employ the topology optimization to design highly-efficient 1D and 2D grating couplers combined with bottom reflectors on the silicon-on-insulator~(SOI) platform and address several key requirements,
such as high coupling efficiency, ultracompact spatial footprints, compatibility with standard single-mode fibers and perfectly vertical coupling scheme, at the same time (Fig.~\ref{fig:TOC}).
Both types of the couplers have minimal dimensions of 14~\textmu m~\texttimes~14~\textmu m and operate at a 0\textdegree~fiber coupling angle to reduce the complexity of alignment routine and optical packaging.
By incorporating bottom reflectors to enhance the directionality of the couplers, the simulated coupling efficiency at the central wavelength of 1550 nm is $-0.37$~dB and $-0.46$~dB for the 1D and 2D grating couplers, respectively.
After the design process, the couplers are fabricated on the 220-nm SOI platform and the bottom reflectors are introduced using the benzocyclobutene~(BCB) wafer bonding technique.
In the experimental characterization, the maximum measured coupling efficiency is $-0.92(1)$~dB for the 1D topology-optimized grating coupler and $-0.86(13)$~dB for the 2D one.
Our experimental demonstration showcases the inverse-designed grating couplers achieving sub-decibel coupling efficiency, ultracompact spatial footprints and compatibility with common single-mode fibers, offering a new solution to various photonic applications requiring high efficiency and high-density integration.

\section*{Topology Optimization for Coupler Design}
\begin{figure*}[hbtp]
  \centering
  \includegraphics[width=1\textwidth]{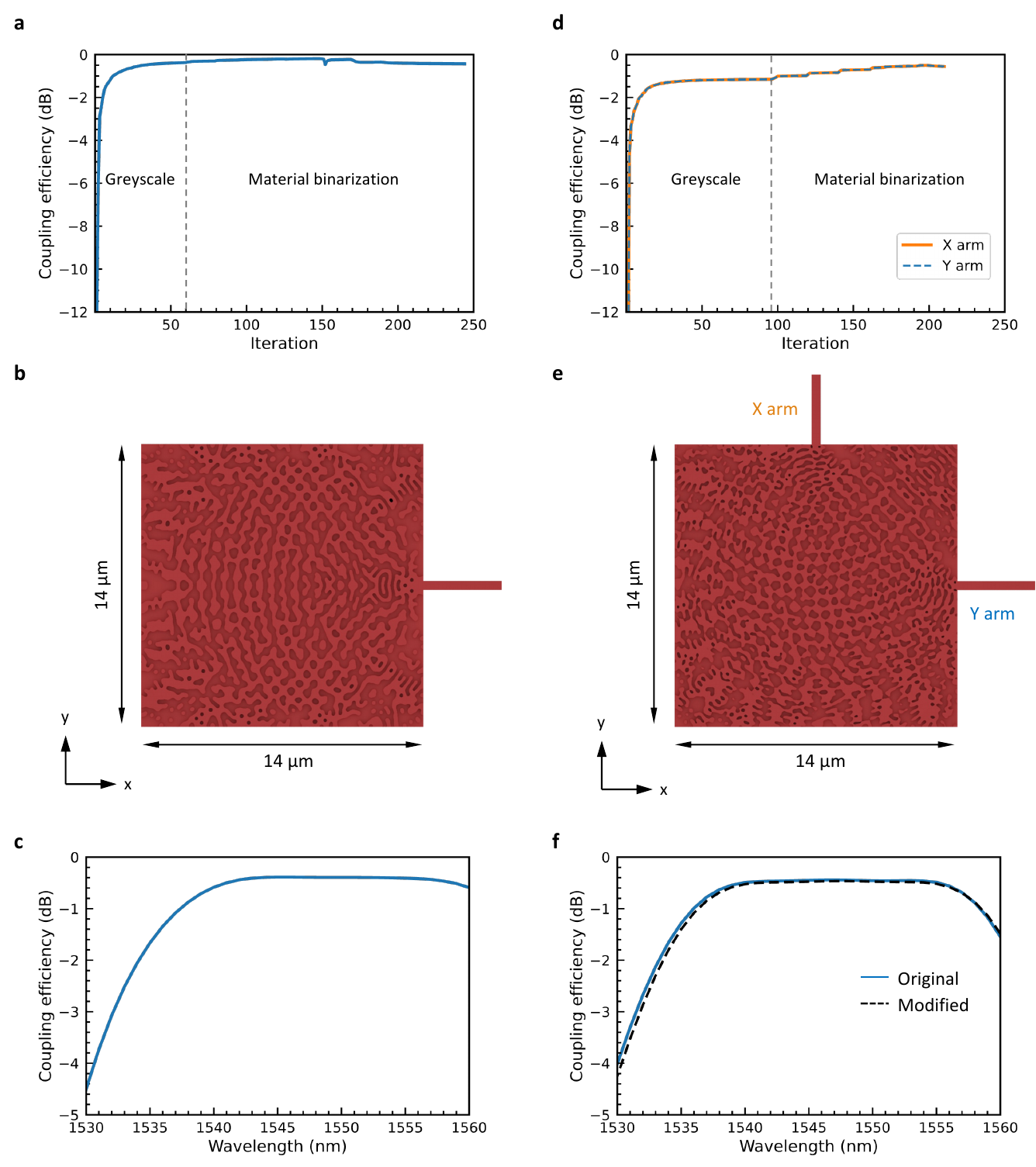}
  \caption{
  Simulation and optimization results of the topology-optimized grating couplers. 
  (a) Evolution of the topology-optimized 1D grating coupler's coupling efficiency during the optimization.
  The material binarization starts after iteration 61.
  (b) Rendered image of the topology-optimized 1D grating coupler.
  (c) Simulated coupling efficiency of the topology-optimized 1D grating coupler within the range from 1530 nm to 1560 nm.
  The coupling efficiency at 1550 nm is $-0.37$ dB.
  (d) Evolution of the topology-optimized 2D grating coupler's coupling efficiency during the optimization.
    The material binarization starts after iteration 97.
  (e) Rendered image of the topology-optimized 2D grating coupler.
  The two waveguides connecting to the 2D grating coupler are labeled as X and Y arms as the light with polarization along $x$ and $y$-axes propagates to the arms, respectively.
  (f) Simulated coupling efficiency of the original (blue solid line) and modified (black dashed line) topology-optimized 2D grating coupler within the range from 1530 nm to 1560 nm.
  The coupling efficiency at 1550 nm is $-0.46$ dB and $-0.47$ dB for the original design and the modified one, respectively.
  }
  \label{fig:final structure}
  \end{figure*}

The topology-optimized couplers with the bottom reflector are designed on the 220-nm SOI platform using the Python-based package LumOpt from Ansys Lumerical, which applies the optimizer based on the L-BFGS-B algorithm.
The coupler structure is discretized in the design region, in which each pixel is assigned a permittivity $\epsilon$ parametrized by $\rho (x,y) \in [0, 1]$:
\begin{align}
  \epsilon(x,y) = \epsilon_{\mathrm{SiO_2}} + \rho(x,y) \cdot ( \epsilon_{\mathrm{Si}} - \epsilon_{\mathrm{SiO_2}}),
\end{align}
where $x$ and $y$ indicate the coordinates of the pixel and $\epsilon_{\mathrm{Si}}$ ($\epsilon_{\mathrm{SiO_2}}$) represents the permittivity of silicon (silicon dioxide).
By addressing the optimization problem, the permittivity of each pixel evolves over iterations, aiming to maximize the user-defined figure of merit (FoM).
The optical field distributions required to calculate the FoM and the gradient of the FoM are computed by performing forward and adjoint simulations using the finite-difference time-domain (FDTD) method \cite{lalau2013adjoint}.
Eventually, the permittivity approaches that of either silicon or silicon dioxide after the threshold parameter of the binarization projection function is ramped up.

All FDTD simulations performed during the topology optimization are three-dimensional to guarantee the integrity of the device performance.
The Gaussian beam source is applied and its waist is set to be 10.4~\textmu m to fit the mode profile of a standard SMF-28 optical fiber positioned on top of the coupler.
Both types of couplers have a minimal design region of 14~\textmu m~\texttimes~14~\textmu m along the $x$ and $y$ directions to entirely accommodate the Gaussian mode of the light emitted from the fiber.
The silicon device layer and the buried oxide layer have a thickness of 220 nm and 2 \textmu m, respectively.
Along the $z$ direction, the design region is 70 nm in height to form the shallow-etched grating structure.
A layer of a perfect electric conductor is placed below the buried oxide layer to serve as the bottom reflector.
The device performance, mainly the coupling efficiency, is defined by the proportion of the optical power coupled into the fundamental transverse electric (TE) mode of the connecting waveguide. 
Specifically, it is the modal overlap integral between the fundamental TE mode of a waveguide with a cross-section of 220 nm~\texttimes~450 nm $E_{\mathrm{wg}}$ and the input waveguide mode $E_{ij}$
\begin{align}
  \mathrm{FoM}_{ij} = \frac{|\int_{S} E_{ij}^* E_{\mathrm{wg}}~dS|^2}{\int_{S}|E_{ij}|^2~dS~\int_{S}|E_{\mathrm{wg}}|^2~dS},
\end{align}
where $i$ indicates the input polarization state and $j$ dictates the index of the output waveguide.
The wavelength range considered for the optimization spans from 1540 nm to 1560 nm with 5 points for constructing a flat optical response within the desired spectral region.
In addition, the minimum feature size of the couplers is implicitly determined by the top-hat filter applied during the topology optimization.
All optimizations are performed on a desktop equipped with an Intel i9-13900K CPU and DDR5 RAM.

In contrast to our previous work \cite{huang2025compact}, the initial condition we choose in this work is the "mixture" one when designing the 1D grating coupler.
This means that a pseudo material, of which the permittivity is the average of silicon and silicon oxide, is filled across the design region ($\rho=0.5$ for all pixels in the first iteration).
The minimum feature size in this case is set to 150 nm, which is slightly higher than the common fabrication limitation.
Given the symmetry of the input polarization state (i.e. the $y$-polarized fiber guided mode LP$^y_{01}$) and the device structure along the $x$ direction, the symmetric boundary condition is applied to reduce the runtime of each FDTD simulation by a factor of two.
The whole optimization is completed in approximately 10 days after 246 iterations (Fig.~\ref{fig:final structure}(a)).
As expected, the final structures have a mirror symmetry along the $x$ direction (Fig.~\ref{fig:final structure}(b)).
As also reported from our previous investigation in Supplementary Information \cite{huang2025compact}, the topology-optimized 1D coupler has a flat optical response within the telecom C-band. 
At the wavelength of 1550 nm, the simulated coupling efficiency is $-0.37$ dB (Fig.~\ref{fig:final structure}(c)).

Similar simulation and optimization configurations are applied for designing the 2D grating coupler using topology optimization.
The 2D grating coupler acts as a polarization demultiplexer by coupling two orthogonal fiber guided modes, LP$^x_{01}$ and LP$^y_{01}$, and splitting them into two individual waveguides extending along the $y$ and $x$ directions, respectively.
Based on reciprocity, light transmitted from these two waveguides is in theory converted into these two orthogonal fiber modes correspondingly.
Since the input polarization varies between $x$- and $y$- polarization states during the optimization for achieving the abovementioned functionality, it is not possible to apply $x$- or $y$-symmetric boundary conditions in the FDTD region to reduce the simulation time.
Again, the initial condition of the optimization is set to be the "mixture" one and a minimum feature size of 150 nm is chosen.
Overall, the optimization is completed after 210 iterations in approximately 37 days (Fig.~\ref{fig:final structure}(d)).

The final structure shares certain similarities with the topology-optimized 1D grating coupler  (Fig.~\ref{fig:final structure}(e)).
Despite the lack of any symmetric boundary conditions, there is a mirror symmetry along the diagonal of the grating structure. 
The whole structure is composed by irregular islands and holes that may reshape the wavefront and guide the light from the fiber to the waveguide and vice versa.
Two groups of confocal rims oriented along the $x$ and $y$ directions seemingly intertwine with each other, resulting in lattice-like patterns in the center of the coupler.
Such a configuration is akin to overlapping two focusing 1D grating couplers and the light is therefore routed to the respective waveguides when the optical modes are gradually converted into the fundamental TE mode of the waveguide.
After this final design is acquired, one additional 3D FDTD simulation with a broadband input light source is also conducted to investigate the device performance. 
The simulated coupling efficiency also has a flat band within the telecom C-band and is $-0.46$ dB at the wavelength of 1550 nm (blue solid line in Fig.~\ref{fig:final structure}(f)).
In addition, to ease the fabrication difficulty, some small holes and islands are manually removed from the final structure (details are presented in Fig.~S1 in Supplementary Information).
Despite this modification, the simulated coupling efficiency matches that of the original design closely and is $-0.47$ dB at the target wavelength of 1550 nm (black dashed line in Fig.~\ref{fig:final structure}(f)).

\section*{Experimental characterization}
\begin{figure*}[hbtp]
  \centering
  \includegraphics[width=1\textwidth]{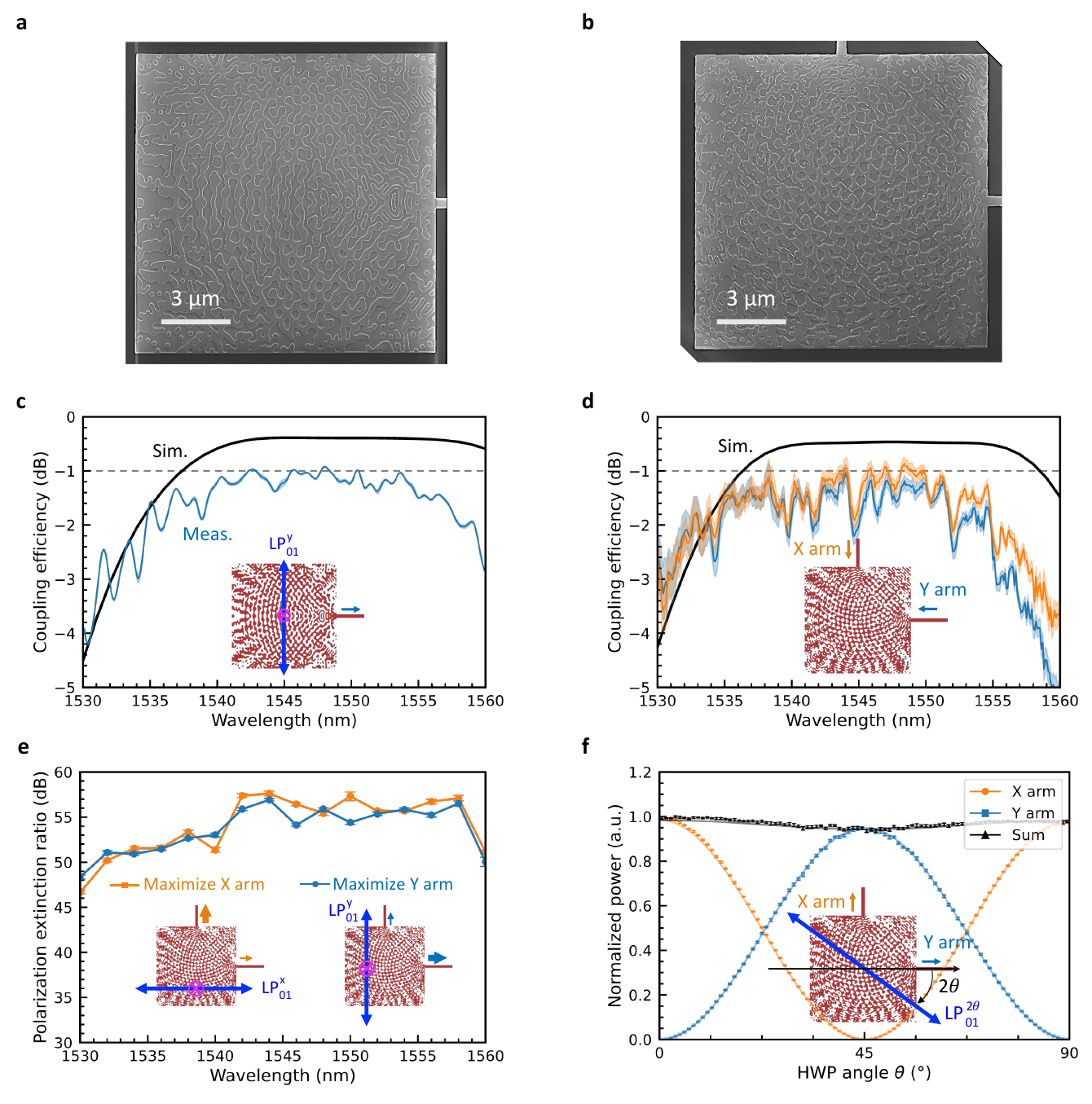}
  \caption{
  Fabricated topology-optimized 1D and 2D grating couplers and the experimental characterization.
  (a) Scanning electron microscope (SEM) image of the topology-optimized 1D grating coupler.
  (b) SEM image of the topology-optimized 2D grating coupler.
  (c) Simulated (black line) and measured (blue line) coupling efficiency of the topology-optimized 1D grating coupler versus wavelength.
  The blue shaded region indicates the deviations of the measurements.
  The measured peak coupling efficiency is $-0.92(1)$ dB at 1548.0 nm and the measured coupling efficiency at 1550.0 nm is $ -1.21(2)$ dB.
  (d) Simulated (black line) and measured coupling efficiency of the topology-optimized 2D grating coupler versus wavelength when the input light is at X (orange line) and Y (blue line) arms.
  The shaded regions indicate the deviations of the measurements.
  The measured peak coupling efficiency is $-0.86(13)$ dB at 1548.4 nm for the light from the X arm and $-1.03(16)$ dB at 1544.0 nm for the light from the Y arm.
  At the target wavelength of 1550 nm, the measured coupling efficiency is $-1.09(10)$ dB and $-1.25(11)$ dB for the light from the X and Y arms, respectively.
  Note that the simulated coupling efficiency shown here corresponds to the post-processed structure.
  (e) Measured polarization extinction ratio of the topology-optimized 2D grating coupler between two arms versus wavelength.
  At the wavelength of 1550 nm, the measured polarization extinction ratio is $57.31(50)$ dB and $54.41(22)$ dB for maximizing the power from  X and Y arms, respectively.
  (f) Measured power dependency of the X (orange) and Y arms (blue), along with the sum of their powers (black), on the HWP angle $\theta$ at the wavelength of 1550 nm.
  }
  \label{fig:experimental characterization}
  \end{figure*}

Both of the couplers are fabricated through CMOS fabrication process (Fig.~\ref{fig:experimental characterization}(a) and (b) and details in Method).
In the case of the topology-optimized 1D grating coupler, two copies are connected back to back by a single waveguide, forming a loop for characterization.
The light is sent into one coupler and coupled out from the other one through a single-mode polarization-maintaining (PM) fiber array with a 0\textdegree~coupling angle (details about the setup are shown in Fig.~S2(a) in Supplementary Information).
The coupling efficiency is then investigated by sweeping the wavelength of the input light from 1530 nm to 1560 nm (Fig.~\ref{fig:experimental characterization}(c)).
To include the fluctuation potentially caused by, for instance, mechanical and thermal instability, we present the mean values (blue line in Fig.~\ref{fig:experimental characterization}(c)) and the standard deviations (blue shaded region in Fig.~\ref{fig:experimental characterization}(c)) of fifty wavelength sweeps, which requires in total approximately 1 hour.
Finally, the coupling efficiency of one topology-optimized 1D coupler is estimated by assuming that two copies of the couplers have the same device performance.
The designed 1D coupler has a peak measured coupling efficiency of $-0.92(1)$ dB at the wavelength of 1548.0~nm.
At the target wavelength of 1550.0~nm, the measured coupling efficiency is $-1.21(2)$ dB, which is $0.84$ dB lower than the simulated value.
This discrepancy between the simulated and measured coupling efficiency could be attributed to the displacement of the grating pattern along the $x$ and $y$ directions (Fig.~S3(a)) together with other fabrication imperfections.

Similarly, the performance of the topology-optimized 2D grating coupler is investigated across the wavelength range from 1530 nm to 1560 nm (Fig.~\ref{fig:experimental characterization}(d)).
In the characterization circuit, the two connecting waveguides of the topology-optimized 2D grating coupler, referred to as X and Y arms, link to conventional 1D grating couplers of which the fiber coupling angle is 14\textdegree~(Fig.~S2(b)).
Two types of fiber arrays are therefore used for coupling light in different conditions:
A 14\textdegree-angled~PM fiber array is positioned on top of the conventional 1D grating couplers to fulfill the requirement of oblique coupling.
At the other side of the circuit, a simple single-mode fiber array is applied to maintain the arbitrary polarization state of the light going into or coming out from the 2D grating coupler.
Again, we acquire the mean values (solid lines in Fig.~\ref{fig:experimental characterization}(d)) and the standard deviations (shaded regions in Fig.~\ref{fig:experimental characterization}(d)) from 50 wavelength sweeps and the total data acquisition time is roughly 1 hour.
When the input light is sent from the X and Y arms to the 2D grating coupler separately, the measured peak coupling efficiency is $-0.86(13)$ dB at 1548.4 nm and $-1.03(16)$ dB at 1544.0 nm, respectively.
At the target wavelength of 1550 nm, the measured coupling efficiency is $-1.09(10)$ dB and $-1.25(11)$ dB for the light from the X and Y arms, respectively.
It is obvious that the measured coupling efficiency differs when the input light is sent from different arms and this imbalance is mostly likely attributed to the grating-pattern displacement and other potential fabrication errors (Fig.~S3(b)).

We also investigate the crosstalk between the two arms that occurs when the 2D grating coupler couples light with only one of the orthogonal polarization states, as this effect degrades the signal fidelity in real applications.
Here, light is injected into the 2D grating coupler and the power from the X or Y arm is maximized by adjusting a manual fiber polarization controller (Fig.~S2(c)). 
The polarization extinction ratio is then calculated in both cases with 16 equally spaced points across 1530 nm to 1560 nm (Fig.~\ref{fig:experimental characterization}(e)).
At each wavelength point, the optical power from the two ports is sampled every 0.1 second over a duration of 5 seconds, resulting in 50 measurements per point.
The lowest measured polarization extinction ratio is $46.66(15)$ dB at the wavelength of 1530 nm.
At the wavelength of 1550 nm, the measured polarization extinction ratios are $57.31(50)$ dB and $54.41(22)$ dB for the cases of maximizing X and Y arms, respectively.

In addition to the high polarization extinction ratio, it is also important to examine the coupler's capability to decompose an arbitrary photonic signal into two orthogonal polarization states and route the light into respective arms with the desired proportion. 
We thereby rotate the half-wave plate (HWP) in the free-space section of the input optical path (Fig.~S2(c)) to change the input polarization state and measure the power from the two arms of the designed 2D grating coupler.
Here, the HWP angle $\theta$ is swept in 1\textdegree~steps over a range from 0\textdegree~to 90\textdegree, yielding in total 91 points (Fig.~\ref{fig:experimental characterization}(f)).
At each measurement point, the optical power from the X and Y arms is sampled every 0.1 seconds for 1 second.
Note that most of the error bars are smaller than the plotted symbols and are therefore barely visible in the figure.

At the wavelength of 1550 nm, the measurement result and the fitted curves show that the maxima of the X-arm curve (orange line in Fig.~\ref{fig:experimental characterization}(f)) coincide the minima of the Y-arm curve (blue line in Fig.~\ref{fig:experimental characterization}(f)), and vice versa.
These two curves are essentially two sinusoidal waves that are in antiphase with slightly different amplitudes and initial phases (see Table.~S1 in Supplementary Information for all fitted parameters).
Specifically, the period of the fitted curves for the X and Y arms is about 90\textdegree.
This indicates that the polarization states that maximize the power in the X and Y arms are orthogonal to each other, as the input polarization state becomes its orthogonal counterpart after the HWP is rotated by 45\textdegree.
Besides, the fitted sinusoidal curve of the sum of the power from the two arms has an offset of about 0.96 with an amplitude of about 0.02 (black curve in Fig.~\ref{fig:experimental characterization}(f)), which corresponds to the discrepancy in the coupling efficiency of X and Y arms observed previously (Fig.~\ref{fig:experimental characterization}(d)).

\begin{table*}[hbtp]
  \renewcommand{\arraystretch}{2.5}
    \centering
    \scalebox{0.85}{
    \begin{tabular}
    {|| c c c c c c c c||}

    \hline
		Year & Ref. 								                & Structure features 									                    & Bottom reflector            & Sim. CE \newline (dB) & Meas. CE \newline (dB) & Designed fiber coupling angle   & Footprint \newline (\textmu m$^2$)\\  [1ex]
		\hline\hline
		
		2014 & [\citenum{zaoui2014bridging}]& \parbox[m]{10em}{Aperiodic gratings with taper waveguide}		                      & \checkmark                  & $-0.33$ 				        & $-0.62^\ddagger$                 & 9\textdegree	          	       	                       & 200$^\dagger$ $\times$ 14 \\ 
		\hline
		
    \multirow{2}{*}{2020} & \multirow{2}{*}{[\citenum{hoppe2020ultra}]}		 		    & \parbox[m]{10em}{Aperiodic gratings with taper waveguide}        & \checkmark      & $-0.33$				        & $-0.50^\star$                 & 9\textdegree	         	                             & 460 $\times$ 15 \\		    
    &              	 		                        & \parbox[m]{10em}{Aperiodic focusing gratings}                                          & \checkmark                  & $-0.75$				        & $-0.93^\star$                 & 9\textdegree	          	       	                       & 30 $\times$ 30 \\		
    \hline
		2022 & [\citenum{liu2022high}]				      & \parbox[m]{10em}{Shallow-etched gratings with taper waveguide}                  & \checkmark                  & $-0.75$				        & $-0.8$                 & 0\textdegree	         	                              & 200$^\dagger$ $\times$ 14 \\
		\hline

    2022 & [\citenum{hammond2022multi}]		      & \parbox[m]{10em}{Multilayer and topology-optimized}                       & \texttimes                  & $-3.0$				        & $-4.7$                 & 0\textdegree	          	         	                     & 10  $\times$ 10   \\ 
		\hline
		2024 & [\citenum{wang2024single}]		     	  & \parbox[m]{10em}{Fully-etched and topology-optimized}                     & \checkmark                  & $-1.08$				        & $-1.44$                & 0\textdegree	                	                       & 2.8  $\times$ 2.8$^*$ \\
		\hline
    2025 & [\citenum{pita2025inverse}]		     	& \parbox[m]{10em}{Shallow-etched and topology-optimized}                   & \texttimes                  & $-2.3$				        & $-3.3$                 & 8\textdegree	         	        	                       & 12  $\times$ 12 \\
		\hline
    2025 & This work 						                & \parbox[m]{10em}{Shallow-etched and topology-optimized}                   & \checkmark                  & $-0.37$				      & $-0.92$               & 0\textdegree	         	        	                       & 14  $\times$ 14  \\ [1ex] 
		\hline
    \end{tabular}
    }
    \caption{Comparison of state-of-the-art 1D grating couplers on the SOI platform for the telecom C-band. 
    Sim./Meas.~CE: simulated/measured coupling efficiency. 
  $^\ddagger$Measured at the fiber coupling angle of 11\textdegree.
  $^\star$Measured at the fiber coupling angle of 10\textdegree.
  $^\dagger$Estimation from given information in the reference.
  $^*$Lens fibers with a mode field diameter of 2.5 \textmu m are applied.
	}
    \label{table: 1D coupler comparison}
\end{table*}

\begin{table*}[hbtp]
  \renewcommand{\arraystretch}{2.5}
    \centering
    \scalebox{0.82}{
    \begin{tabular}
    {|| c c c c c c c c||}
    \hline
		Year & Ref. 								                & Structure features 									                    & Bottom reflector            & Sim. CE \newline (dB) & Meas. CE \newline (dB) & Designed fiber coupling angle   & Footprint \newline of one arm \newline (\textmu m$^2$)\\  [1ex]
		\hline\hline
		2018 & [\citenum{luo2018low}]& \parbox[m]{10em}{Shallow-etched hole array with taper waveguides}		                      & \checkmark                  & $-1.37$ 				        & $-1.80$               & 12\textdegree	         	       	                       & 600$^\dagger$ $\times$ 11 \\ 
		\hline
		2019 & [\citenum{watanabe20192}]& \parbox[m]{10em}{Overlapped blazed gratings with taper waveguides}		                      & \texttimes                  & $-2.4$ 				        & $-2.6$                 & 0\textdegree	         	       	                       & 300 $\times$ 10$^\dagger$ \\ 
		\hline
		
    2025 & [\citenum{zhou2025efficient}]		      & \parbox[m]{10em}{Shallow-etched hole array and poly-Si cylinder array with taper waveguides}                       & \texttimes                  & $-2.37$				        & $-2.54$                 & 0\textdegree	         	         	                     & 370$^\dagger$  $\times$ 13.2   \\ 
		\hline

    \multirow{2}{*}{2025} & \multirow{2}{*}{[\citenum{luo2025silicon}]}		 		    & \multirow{2}{*}{\shortstack[c]{Fully-etched \\ inverse-designed gratings \\ with taper waveguides}}	        & \texttimes      & $-3.2$				        & $-5.3$                 & \multirow{2}{*}{0\textdegree}	 	                                      &   \multirow{2}{*}{500$^\dagger$ $\times$ 14} \\		    
                        &              	 		                        &                                                                      & \checkmark     & $-1.9$				        & $-2.9$                 &        	       	                       & \\		
    
    \hline

    2022 & [\citenum{hammond2022multi}]		     	& \parbox[m]{10em}{Multilayer and topology-optimized}                   & \texttimes                  & $-5.6$				        & $-7.0$                 & 0\textdegree	         	        	                       & 10  $\times$ 10 \\
		\hline
    2025 & [\citenum{wu2025ultra}]		     	& \parbox[m]{10em}{Fully-etched and topology-optimized}                   & \checkmark                  & $-0.98$				        & $-1.05$                 & 0\textdegree	         	        	                       & 2.8  $\times$ 2.8$^*$ \\
		\hline
    2025 & This work 						                & \parbox[m]{10em}{Shallow-etched and topology-optimized}                   & \checkmark                  & $-0.46$				      & $-0.86$               & 0\textdegree	         	        	                       & 14  $\times$ 14  \\ [1ex] 
		\hline
    \end{tabular}
    }
    \caption{Comparison of state-of-the-art 2D grating couplers on the SOI platform for the telecom C-band. 
    Sim./Meas.~CE: simulated/measured coupling efficiency. 
  $^\dagger$Estimation from given information in the reference.
  $^*$Lens fibers with a mode field diameter of 2.5 \textmu m are applied.
	}
    \label{table: 2D coupler comparison}
\end{table*}

\section*{Conclusion}
In summary, we design and experimentally demonstrate the topology-optimized 1D and 2D couplers featuring ultracompact footprint, a perfectly vertical coupling scheme, compatibility with standard single-mode fibers and sub-decibel coupling efficiency.
Both types of couplers have dimensions of only 14~\textmu m~\texttimes~14~\textmu m on the SOI platform and operate within the telecom C-band with 0\textdegree~fiber coupling angle.
For the topology-optimized 1D grating coupler, the simulated coupling efficiency is $-0.37$ dB at the central wavelength of 1550 nm and the measured peak coupling efficiency is $-0.92(1)$ dB at the wavelength of 1548.0 nm.
As for the 2D grating coupler, the simulated coupling efficiency is $-0.46$ dB at 1550 nm and the measured peak coupling efficiency is $-0.86(13)$ dB at 1548.4 nm.
In addition, the measured polarization extinction ratios are 57.31(50) dB and 54.41(22) dB at 1550.0 nm, respectively.
Based on the experimental characterization, our ultracompact design of the 2D grating coupler, to the best of our knowledge, attains the lowest insertion loss to date in the literature (Table.~\ref{table: 2D coupler comparison}).

These miniaturized couplers are advantageous for constructing large-scale photonic integrated systems requiring both high efficiency and high-density integration.
For instance, interfacing the photonic integrated circuits applying space-division multiplexing using multi-core fibers for increasing the capacity of optical data transmission \cite{liu2022high}.
Moreover, in combination with on-chip Mach-Zehnder interferometers, which may also be composed of topology-optimized 2\texttimes2 beam splitters, and thermo-optic phase shifters, these couplers can further downsize the entire photonic integrated circuits for applications such as quantum state distribution with conversion between the path encoding and polarization encoding~\cite{wang2016chip,liu2022generation}, quantum key distribution~\cite{zhang2019integrated} and quantum teleportation~\cite{llewellyn2020chip}.

\section*{Methods}
\subsection*{Device fabrication}
\vspace{-1em}
Both of the topology-optimized 1D and 2D couplers are fabricated by the commercial foundry through a dedicated run (Fig.~\ref{fig:experimental characterization}(a) and (b)). 
The couplers are created on a 220-nm silicon device layer.
Specifically, the grating structures are patterned through a two-step etching process.
A 2-\textmu m bottom oxide layer is underneath the device layer and a 2-\textmu m top oxide layer is later deposited as a protection layer.
The BCB bonding technique is applied to produce the aluminum layer that serves as the bottom reflector across the whole wafer.

\subsection*{Experimental setup}
\vspace{-1em}
For the experimental characterization (Fig.~S2 in Supplementary Information), a tunable continuous-wave laser source is used for investigating the device performance.
The light is coupled into and from the couplers via standard single-mode fiber arrays positioned by photonic alignment systems.
The resulting optical power is measured using a high-resolution photometer.

\section*{Data availability}
\vspace{-0.8em}
Data presented in this study and supporting the findings are available from the corresponding author upon reasonable request.

\section*{Fundings}
\vspace{-0.8em}
The authors acknowledge the funding support from the Carl Zeiss Foundation, the Centre for Integrated Quantum Science and Technology (IQST), 
the Federal Ministry of Research, Technology and Space (BMFTR, projects SiSiQ: FKZ 13N14920, PhotonQ: FKZ 13N15758, QRN: FKZ16KIS2207), 
the Ministry of Science, Research and Arts Baden-Württemberg, and the Deutsche Forschungsgemeinschaft (DFG, German Research Foundation, 431314977/GRK2642, SFB 1667).

\section*{Contributions}
\vspace{-0.8em}
S.B. and S.-Y.H. conceived the idea and designed the study. S.-Y.H. performed the optimization and numerical simulations of the couplers. 
J.H. and J.Z. designed the integrated circuits. 
S.-Y.H. carried out the experimental characterization and data analysis, with constructive discussions with J.Z., T.E. and J.H. 
S.-Y.H. wrote the manuscript with input and feedback from all authors.

\section*{Corresponding authors}
\vspace{-0.8em}
Correspondence to Shiang-Yu Huang.

\section*{Competing interests}
\vspace{-0.8em}
The authors declare no competing interests.

\bibliography{achemso-demo-jabbrv}

\begin{thebibliography}{21}%
\makeatletter
\providecommand \@ifxundefined [1]{%
 \@ifx{#1\undefined}
}%
\providecommand \@ifnum [1]{%
 \ifnum #1\expandafter \@firstoftwo
 \else \expandafter \@secondoftwo
 \fi
}%
\providecommand \@ifx [1]{%
 \ifx #1\expandafter \@firstoftwo
 \else \expandafter \@secondoftwo
 \fi
}%
\providecommand \natexlab [1]{#1}%
\providecommand \enquote  [1]{``#1''}%
\providecommand \bibnamefont  [1]{#1}%
\providecommand \bibfnamefont [1]{#1}%
\providecommand \citenamefont [1]{#1}%
\providecommand \href@noop [0]{\@secondoftwo}%
\providecommand \href [0]{\begingroup \@sanitize@url \@href}%
\providecommand \@href[1]{\@@startlink{#1}\@@href}%
\providecommand \@@href[1]{\endgroup#1\@@endlink}%
\providecommand \@sanitize@url [0]{\catcode `\\12\catcode `\$12\catcode `\&12\catcode `\#12\catcode `\^12\catcode `\_12\catcode `\%12\relax}%
\providecommand \@@startlink[1]{}%
\providecommand \@@endlink[0]{}%
\providecommand \url  [0]{\begingroup\@sanitize@url \@url }%
\providecommand \@url [1]{\endgroup\@href {#1}{\urlprefix }}%
\providecommand \urlprefix  [0]{URL }%
\providecommand \Eprint [0]{\href }%
\providecommand \doibase [0]{https://doi.org/}%
\providecommand \selectlanguage [0]{\@gobble}%
\providecommand \bibinfo  [0]{\@secondoftwo}%
\providecommand \bibfield  [0]{\@secondoftwo}%
\providecommand \translation [1]{[#1]}%
\providecommand \BibitemOpen [0]{}%
\providecommand \bibitemStop [0]{}%
\providecommand \bibitemNoStop [0]{.\EOS\space}%
\providecommand \EOS [0]{\spacefactor3000\relax}%
\providecommand \BibitemShut  [1]{\csname bibitem#1\endcsname}%
\let\auto@bib@innerbib\@empty
\bibitem [{\citenamefont {Hatori}\ \emph {et~al.}(2014)\citenamefont {Hatori}, \citenamefont {Shimizu}, \citenamefont {Okano}, \citenamefont {Ishizaka}, \citenamefont {Yamamoto}, \citenamefont {Urino}, \citenamefont {Mori}, \citenamefont {Nakamura},\ and\ \citenamefont {Arakawa}}]{hatori2014hybrid}%
  \BibitemOpen
  \bibfield  {author} {\bibinfo {author} {\bibfnamefont {N.}~\bibnamefont {Hatori}}, \bibinfo {author} {\bibfnamefont {T.}~\bibnamefont {Shimizu}}, \bibinfo {author} {\bibfnamefont {M.}~\bibnamefont {Okano}}, \bibinfo {author} {\bibfnamefont {M.}~\bibnamefont {Ishizaka}}, \bibinfo {author} {\bibfnamefont {T.}~\bibnamefont {Yamamoto}}, \bibinfo {author} {\bibfnamefont {Y.}~\bibnamefont {Urino}}, \bibinfo {author} {\bibfnamefont {M.}~\bibnamefont {Mori}}, \bibinfo {author} {\bibfnamefont {T.}~\bibnamefont {Nakamura}},\ and\ \bibinfo {author} {\bibfnamefont {Y.}~\bibnamefont {Arakawa}},\ }\bibfield  {title} {\bibinfo {title} {A hybrid integrated light source on a silicon platform using a trident spot-size converter},\ }\href@noop {} {\bibfield  {journal} {\bibinfo  {journal} {J. Light. Technol.}\ }\textbf {\bibinfo {volume} {32}},\ \bibinfo {pages} {1329} (\bibinfo {year} {2014})}\BibitemShut {NoStop}%
\bibitem [{\citenamefont {He}\ \emph {et~al.}(2020)\citenamefont {He}, \citenamefont {Guo}, \citenamefont {Wang}, \citenamefont {Zhang},\ and\ \citenamefont {Su}}]{he2020low}%
  \BibitemOpen
  \bibfield  {author} {\bibinfo {author} {\bibfnamefont {A.}~\bibnamefont {He}}, \bibinfo {author} {\bibfnamefont {X.}~\bibnamefont {Guo}}, \bibinfo {author} {\bibfnamefont {K.}~\bibnamefont {Wang}}, \bibinfo {author} {\bibfnamefont {Y.}~\bibnamefont {Zhang}},\ and\ \bibinfo {author} {\bibfnamefont {Y.}~\bibnamefont {Su}},\ }\bibfield  {title} {\bibinfo {title} {Low loss, large bandwidth fiber-chip edge couplers based on silicon-on-insulator platform},\ }\href@noop {} {\bibfield  {journal} {\bibinfo  {journal} {J. Light. Technol.}\ }\textbf {\bibinfo {volume} {38}},\ \bibinfo {pages} {4780} (\bibinfo {year} {2020})}\BibitemShut {NoStop}%
\bibitem [{\citenamefont {Mu}\ \emph {et~al.}(2020)\citenamefont {Mu}, \citenamefont {Wu}, \citenamefont {Cheng},\ and\ \citenamefont {Fu}}]{mu2020edge}%
  \BibitemOpen
  \bibfield  {author} {\bibinfo {author} {\bibfnamefont {X.}~\bibnamefont {Mu}}, \bibinfo {author} {\bibfnamefont {S.}~\bibnamefont {Wu}}, \bibinfo {author} {\bibfnamefont {L.}~\bibnamefont {Cheng}},\ and\ \bibinfo {author} {\bibfnamefont {H.}~\bibnamefont {Fu}},\ }\bibfield  {title} {\bibinfo {title} {Edge couplers in silicon photonic integrated circuits: A review},\ }\href@noop {} {\bibfield  {journal} {\bibinfo  {journal} {Appl. Sci.}\ }\textbf {\bibinfo {volume} {10}},\ \bibinfo {pages} {1538} (\bibinfo {year} {2020})}\BibitemShut {NoStop}%
\bibitem [{\citenamefont {Zaoui}\ \emph {et~al.}(2014)\citenamefont {Zaoui}, \citenamefont {Kunze}, \citenamefont {Vogel}, \citenamefont {Berroth}, \citenamefont {Butschke}, \citenamefont {Letzkus},\ and\ \citenamefont {Burghartz}}]{zaoui2014bridging}%
  \BibitemOpen
  \bibfield  {author} {\bibinfo {author} {\bibfnamefont {W.~S.}\ \bibnamefont {Zaoui}}, \bibinfo {author} {\bibfnamefont {A.}~\bibnamefont {Kunze}}, \bibinfo {author} {\bibfnamefont {W.}~\bibnamefont {Vogel}}, \bibinfo {author} {\bibfnamefont {M.}~\bibnamefont {Berroth}}, \bibinfo {author} {\bibfnamefont {J.}~\bibnamefont {Butschke}}, \bibinfo {author} {\bibfnamefont {F.}~\bibnamefont {Letzkus}},\ and\ \bibinfo {author} {\bibfnamefont {J.}~\bibnamefont {Burghartz}},\ }\bibfield  {title} {\bibinfo {title} {Bridging the gap between optical fibers and silicon photonic integrated circuits},\ }\href@noop {} {\bibfield  {journal} {\bibinfo  {journal} {Opt. Express}\ }\textbf {\bibinfo {volume} {22}},\ \bibinfo {pages} {1277} (\bibinfo {year} {2014})}\BibitemShut {NoStop}%
\bibitem [{\citenamefont {Hoppe}\ \emph {et~al.}(2020)\citenamefont {Hoppe}, \citenamefont {Zaoui}, \citenamefont {Rathgeber}, \citenamefont {Wang}, \citenamefont {Klenk}, \citenamefont {Vogel}, \citenamefont {Kaschel}, \citenamefont {Portalupi}, \citenamefont {Burghartz},\ and\ \citenamefont {Berroth}}]{hoppe2020ultra}%
  \BibitemOpen
  \bibfield  {author} {\bibinfo {author} {\bibfnamefont {N.}~\bibnamefont {Hoppe}}, \bibinfo {author} {\bibfnamefont {W.~S.}\ \bibnamefont {Zaoui}}, \bibinfo {author} {\bibfnamefont {L.}~\bibnamefont {Rathgeber}}, \bibinfo {author} {\bibfnamefont {Y.}~\bibnamefont {Wang}}, \bibinfo {author} {\bibfnamefont {R.~H.}\ \bibnamefont {Klenk}}, \bibinfo {author} {\bibfnamefont {W.}~\bibnamefont {Vogel}}, \bibinfo {author} {\bibfnamefont {M.}~\bibnamefont {Kaschel}}, \bibinfo {author} {\bibfnamefont {S.~L.}\ \bibnamefont {Portalupi}}, \bibinfo {author} {\bibfnamefont {J.}~\bibnamefont {Burghartz}},\ and\ \bibinfo {author} {\bibfnamefont {M.}~\bibnamefont {Berroth}},\ }\bibfield  {title} {\bibinfo {title} {Ultra-efficient silicon-on-insulator grating couplers with backside metal mirrors},\ }\href@noop {} {\bibfield  {journal} {\bibinfo  {journal} {IEEE J. Sel. Top. Quantum Electron.}\ }\textbf {\bibinfo {volume} {26}},\ \bibinfo {pages} {1} (\bibinfo {year} {2020})}\BibitemShut {NoStop}%
\bibitem [{\citenamefont {Liu}\ \emph {et~al.}(2022{\natexlab{a}})\citenamefont {Liu}, \citenamefont {Zheng}, \citenamefont {Chen}, \citenamefont {Wang}, \citenamefont {Li}, \citenamefont {Chen},\ and\ \citenamefont {Liu}}]{liu2022high}%
  \BibitemOpen
  \bibfield  {author} {\bibinfo {author} {\bibfnamefont {J.}~\bibnamefont {Liu}}, \bibinfo {author} {\bibfnamefont {Z.}~\bibnamefont {Zheng}}, \bibinfo {author} {\bibfnamefont {B.}~\bibnamefont {Chen}}, \bibinfo {author} {\bibfnamefont {Z.}~\bibnamefont {Wang}}, \bibinfo {author} {\bibfnamefont {C.}~\bibnamefont {Li}}, \bibinfo {author} {\bibfnamefont {K.}~\bibnamefont {Chen}},\ and\ \bibinfo {author} {\bibfnamefont {L.}~\bibnamefont {Liu}},\ }\bibfield  {title} {\bibinfo {title} {High-performance grating coupler array on silicon for a perfectly-vertically mounted multicore fiber},\ }\href@noop {} {\bibfield  {journal} {\bibinfo  {journal} {J. Light. Technol.}\ }\textbf {\bibinfo {volume} {40}},\ \bibinfo {pages} {5654} (\bibinfo {year} {2022}{\natexlab{a}})}\BibitemShut {NoStop}%
\bibitem [{\citenamefont {Luo}\ \emph {et~al.}(2018)\citenamefont {Luo}, \citenamefont {Nong}, \citenamefont {Gao}, \citenamefont {Huang}, \citenamefont {Zhu}, \citenamefont {Liu}, \citenamefont {Zhou}, \citenamefont {Xu}, \citenamefont {Liu}, \citenamefont {Yu} \emph {et~al.}}]{luo2018low}%
  \BibitemOpen
  \bibfield  {author} {\bibinfo {author} {\bibfnamefont {Y.}~\bibnamefont {Luo}}, \bibinfo {author} {\bibfnamefont {Z.}~\bibnamefont {Nong}}, \bibinfo {author} {\bibfnamefont {S.}~\bibnamefont {Gao}}, \bibinfo {author} {\bibfnamefont {H.}~\bibnamefont {Huang}}, \bibinfo {author} {\bibfnamefont {Y.}~\bibnamefont {Zhu}}, \bibinfo {author} {\bibfnamefont {L.}~\bibnamefont {Liu}}, \bibinfo {author} {\bibfnamefont {L.}~\bibnamefont {Zhou}}, \bibinfo {author} {\bibfnamefont {J.}~\bibnamefont {Xu}}, \bibinfo {author} {\bibfnamefont {L.}~\bibnamefont {Liu}}, \bibinfo {author} {\bibfnamefont {S.}~\bibnamefont {Yu}}, \emph {et~al.},\ }\bibfield  {title} {\bibinfo {title} {Low-loss two-dimensional silicon photonic grating coupler with a backside metal mirror},\ }\href@noop {} {\bibfield  {journal} {\bibinfo  {journal} {Opt. Lett.}\ }\textbf {\bibinfo {volume} {43}},\ \bibinfo {pages} {474} (\bibinfo {year} {2018})}\BibitemShut {NoStop}%
\bibitem [{\citenamefont {Watanabe}\ \emph {et~al.}(2019)\citenamefont {Watanabe}, \citenamefont {Fedoryshyn},\ and\ \citenamefont {Leuthold}}]{watanabe20192}%
  \BibitemOpen
  \bibfield  {author} {\bibinfo {author} {\bibfnamefont {T.}~\bibnamefont {Watanabe}}, \bibinfo {author} {\bibfnamefont {Y.}~\bibnamefont {Fedoryshyn}},\ and\ \bibinfo {author} {\bibfnamefont {J.}~\bibnamefont {Leuthold}},\ }\bibfield  {title} {\bibinfo {title} {2-{D} grating couplers for vertical fiber coupling in two polarizations},\ }\href@noop {} {\bibfield  {journal} {\bibinfo  {journal} {IEEE Photonics J.}\ }\textbf {\bibinfo {volume} {11}},\ \bibinfo {pages} {1} (\bibinfo {year} {2019})}\BibitemShut {NoStop}%
\bibitem [{\citenamefont {Zhou}\ \emph {et~al.}(2025)\citenamefont {Zhou}, \citenamefont {Lu}, \citenamefont {Kang}, \citenamefont {Wu},\ and\ \citenamefont {Tong}}]{zhou2025efficient}%
  \BibitemOpen
  \bibfield  {author} {\bibinfo {author} {\bibfnamefont {W.}~\bibnamefont {Zhou}}, \bibinfo {author} {\bibfnamefont {K.}~\bibnamefont {Lu}}, \bibinfo {author} {\bibfnamefont {S.}~\bibnamefont {Kang}}, \bibinfo {author} {\bibfnamefont {X.}~\bibnamefont {Wu}},\ and\ \bibinfo {author} {\bibfnamefont {Y.}~\bibnamefont {Tong}},\ }\bibfield  {title} {\bibinfo {title} {Efficient polarization-diversity grating coupler with multipolar radiation mode enhancement},\ }\href@noop {} {\bibfield  {journal} {\bibinfo  {journal} {IEEE Photonics J.}\ } (\bibinfo {year} {2025})}\BibitemShut {NoStop}%
\bibitem [{\citenamefont {Luo}\ \emph {et~al.}(2025)\citenamefont {Luo}, \citenamefont {Wu}, \citenamefont {Chen}, \citenamefont {Yu}, \citenamefont {Chen}, \citenamefont {Gao},\ and\ \citenamefont {Ge}}]{luo2025silicon}%
  \BibitemOpen
  \bibfield  {author} {\bibinfo {author} {\bibfnamefont {Y.}~\bibnamefont {Luo}}, \bibinfo {author} {\bibfnamefont {M.}~\bibnamefont {Wu}}, \bibinfo {author} {\bibfnamefont {B.}~\bibnamefont {Chen}}, \bibinfo {author} {\bibfnamefont {S.}~\bibnamefont {Yu}}, \bibinfo {author} {\bibfnamefont {P.}~\bibnamefont {Chen}}, \bibinfo {author} {\bibfnamefont {S.}~\bibnamefont {Gao}},\ and\ \bibinfo {author} {\bibfnamefont {R.}~\bibnamefont {Ge}},\ }\bibfield  {title} {\bibinfo {title} {Silicon perfectly-vertical polarization diversity grating coupler with near-zero polarization dependent loss},\ }\href@noop {} {\bibfield  {journal} {\bibinfo  {journal} {J. Light. Technol.}\ }\textbf {\bibinfo {volume} {43}},\ \bibinfo {pages} {7777} (\bibinfo {year} {2025})}\BibitemShut {NoStop}%
\bibitem [{\citenamefont {Kuster}\ \emph {et~al.}(2025)\citenamefont {Kuster}, \citenamefont {Augenstein}, \citenamefont {Rockstuhl},\ and\ \citenamefont {Sturges}}]{kuster2025three}%
  \BibitemOpen
  \bibfield  {author} {\bibinfo {author} {\bibfnamefont {O.}~\bibnamefont {Kuster}}, \bibinfo {author} {\bibfnamefont {Y.}~\bibnamefont {Augenstein}}, \bibinfo {author} {\bibfnamefont {C.}~\bibnamefont {Rockstuhl}},\ and\ \bibinfo {author} {\bibfnamefont {T.~J.}\ \bibnamefont {Sturges}},\ }\bibfield  {title} {\bibinfo {title} {A three-dimensional polarization-insensitive grating coupler tailored for 3d nanoprinting},\ }\href@noop {} {\bibfield  {journal} {\bibinfo  {journal} {arXiv preprint arXiv:2508.20894}\ } (\bibinfo {year} {2025})}\BibitemShut {NoStop}%
\bibitem [{\citenamefont {Hammond}\ \emph {et~al.}(2022)\citenamefont {Hammond}, \citenamefont {Slaby}, \citenamefont {Probst},\ and\ \citenamefont {Ralph}}]{hammond2022multi}%
  \BibitemOpen
  \bibfield  {author} {\bibinfo {author} {\bibfnamefont {A.~M.}\ \bibnamefont {Hammond}}, \bibinfo {author} {\bibfnamefont {J.~B.}\ \bibnamefont {Slaby}}, \bibinfo {author} {\bibfnamefont {M.~J.}\ \bibnamefont {Probst}},\ and\ \bibinfo {author} {\bibfnamefont {S.~E.}\ \bibnamefont {Ralph}},\ }\bibfield  {title} {\bibinfo {title} {Multi-layer inverse design of vertical grating couplers for high-density, commercial foundry interconnects},\ }\href@noop {} {\bibfield  {journal} {\bibinfo  {journal} {Opt. Express}\ }\textbf {\bibinfo {volume} {30}},\ \bibinfo {pages} {31058} (\bibinfo {year} {2022})}\BibitemShut {NoStop}%
\bibitem [{\citenamefont {Wang}\ \emph {et~al.}(2024)\citenamefont {Wang}, \citenamefont {Qiu}, \citenamefont {Dong}, \citenamefont {Chen}, \citenamefont {Wu}, \citenamefont {Guo},\ and\ \citenamefont {Wu}}]{wang2024single}%
  \BibitemOpen
  \bibfield  {author} {\bibinfo {author} {\bibfnamefont {L.}~\bibnamefont {Wang}}, \bibinfo {author} {\bibfnamefont {J.}~\bibnamefont {Qiu}}, \bibinfo {author} {\bibfnamefont {Z.}~\bibnamefont {Dong}}, \bibinfo {author} {\bibfnamefont {Y.}~\bibnamefont {Chen}}, \bibinfo {author} {\bibfnamefont {L.}~\bibnamefont {Wu}}, \bibinfo {author} {\bibfnamefont {H.}~\bibnamefont {Guo}},\ and\ \bibinfo {author} {\bibfnamefont {J.}~\bibnamefont {Wu}},\ }\bibfield  {title} {\bibinfo {title} {Single-etched fiber-chip coupler with a metal mirror on a 220-nm silicon-on-insulator platform for perfectly vertical coupling},\ }\href@noop {} {\bibfield  {journal} {\bibinfo  {journal} {Opt. Lett.}\ }\textbf {\bibinfo {volume} {49}},\ \bibinfo {pages} {2974} (\bibinfo {year} {2024})}\BibitemShut {NoStop}%
\bibitem [{\citenamefont {Pita}\ \emph {et~al.}(2025)\citenamefont {Pita}, \citenamefont {Dainese},\ and\ \citenamefont {M{\'e}nard}}]{pita2025inverse}%
  \BibitemOpen
  \bibfield  {author} {\bibinfo {author} {\bibfnamefont {J.}~\bibnamefont {Pita}}, \bibinfo {author} {\bibfnamefont {P.}~\bibnamefont {Dainese}},\ and\ \bibinfo {author} {\bibfnamefont {M.}~\bibnamefont {M{\'e}nard}},\ }\bibfield  {title} {\bibinfo {title} {Inverse design fiber-to-chip couplers for the {O} -and {C}-bands},\ }\href@noop {} {\bibfield  {journal} {\bibinfo  {journal} {Opt. Lett.}\ }\textbf {\bibinfo {volume} {50}},\ \bibinfo {pages} {1973} (\bibinfo {year} {2025})}\BibitemShut {NoStop}%
\bibitem [{\citenamefont {Wu}\ \emph {et~al.}(2025)\citenamefont {Wu}, \citenamefont {Qiu}, \citenamefont {Wang}, \citenamefont {Chen}, \citenamefont {Guo}, \citenamefont {Wei},\ and\ \citenamefont {Wu}}]{wu2025ultra}%
  \BibitemOpen
  \bibfield  {author} {\bibinfo {author} {\bibfnamefont {L.}~\bibnamefont {Wu}}, \bibinfo {author} {\bibfnamefont {J.}~\bibnamefont {Qiu}}, \bibinfo {author} {\bibfnamefont {L.}~\bibnamefont {Wang}}, \bibinfo {author} {\bibfnamefont {Y.}~\bibnamefont {Chen}}, \bibinfo {author} {\bibfnamefont {H.}~\bibnamefont {Guo}}, \bibinfo {author} {\bibfnamefont {S.}~\bibnamefont {Wei}},\ and\ \bibinfo {author} {\bibfnamefont {J.}~\bibnamefont {Wu}},\ }\bibfield  {title} {\bibinfo {title} {Ultra-low-loss polarization-splitting grating coupler based on inverse-design},\ }\href@noop {} {\bibfield  {journal} {\bibinfo  {journal} {Opt. Express}\ }\textbf {\bibinfo {volume} {33}},\ \bibinfo {pages} {7620} (\bibinfo {year} {2025})}\BibitemShut {NoStop}%
\bibitem [{\citenamefont {Lalau-Keraly}\ \emph {et~al.}(2013)\citenamefont {Lalau-Keraly}, \citenamefont {Bhargava}, \citenamefont {Miller},\ and\ \citenamefont {Yablonovitch}}]{lalau2013adjoint}%
  \BibitemOpen
  \bibfield  {author} {\bibinfo {author} {\bibfnamefont {C.~M.}\ \bibnamefont {Lalau-Keraly}}, \bibinfo {author} {\bibfnamefont {S.}~\bibnamefont {Bhargava}}, \bibinfo {author} {\bibfnamefont {O.~D.}\ \bibnamefont {Miller}},\ and\ \bibinfo {author} {\bibfnamefont {E.}~\bibnamefont {Yablonovitch}},\ }\bibfield  {title} {\bibinfo {title} {Adjoint shape optimization applied to electromagnetic design},\ }\href@noop {} {\bibfield  {journal} {\bibinfo  {journal} {Opt. Express}\ }\textbf {\bibinfo {volume} {21}},\ \bibinfo {pages} {21693} (\bibinfo {year} {2013})}\BibitemShut {NoStop}%
\bibitem [{\citenamefont {Huang}\ and\ \citenamefont {Barz}(2025)}]{huang2025compact}%
  \BibitemOpen
  \bibfield  {author} {\bibinfo {author} {\bibfnamefont {S.-Y.}\ \bibnamefont {Huang}}\ and\ \bibinfo {author} {\bibfnamefont {S.}~\bibnamefont {Barz}},\ }\bibfield  {title} {\bibinfo {title} {Compact inverse designed vertical coupler with bottom reflector for sub-decibel fiber-to-chip coupling on silicon on insulator platform},\ }\href@noop {} {\bibfield  {journal} {\bibinfo  {journal} {Sci. Rep.}\ }\textbf {\bibinfo {volume} {15}},\ \bibinfo {pages} {2925} (\bibinfo {year} {2025})}\BibitemShut {NoStop}%
\bibitem [{\citenamefont {Wang}\ \emph {et~al.}(2016)\citenamefont {Wang}, \citenamefont {Bonneau}, \citenamefont {Villa}, \citenamefont {Silverstone}, \citenamefont {Santagati}, \citenamefont {Miki}, \citenamefont {Yamashita}, \citenamefont {Fujiwara}, \citenamefont {Sasaki}, \citenamefont {Terai} \emph {et~al.}}]{wang2016chip}%
  \BibitemOpen
  \bibfield  {author} {\bibinfo {author} {\bibfnamefont {J.}~\bibnamefont {Wang}}, \bibinfo {author} {\bibfnamefont {D.}~\bibnamefont {Bonneau}}, \bibinfo {author} {\bibfnamefont {M.}~\bibnamefont {Villa}}, \bibinfo {author} {\bibfnamefont {J.~W.}\ \bibnamefont {Silverstone}}, \bibinfo {author} {\bibfnamefont {R.}~\bibnamefont {Santagati}}, \bibinfo {author} {\bibfnamefont {S.}~\bibnamefont {Miki}}, \bibinfo {author} {\bibfnamefont {T.}~\bibnamefont {Yamashita}}, \bibinfo {author} {\bibfnamefont {M.}~\bibnamefont {Fujiwara}}, \bibinfo {author} {\bibfnamefont {M.}~\bibnamefont {Sasaki}}, \bibinfo {author} {\bibfnamefont {H.}~\bibnamefont {Terai}}, \emph {et~al.},\ }\bibfield  {title} {\bibinfo {title} {Chip-to-chip quantum photonic interconnect by path-polarization interconversion},\ }\href@noop {} {\bibfield  {journal} {\bibinfo  {journal} {Optica}\ }\textbf {\bibinfo {volume} {3}},\ \bibinfo {pages} {407} (\bibinfo {year} {2016})}\BibitemShut {NoStop}%
\bibitem [{\citenamefont {Liu}\ \emph {et~al.}(2022{\natexlab{b}})\citenamefont {Liu}, \citenamefont {Zheng}, \citenamefont {Yu}, \citenamefont {Feng}, \citenamefont {Liu}, \citenamefont {Cui}, \citenamefont {Huang},\ and\ \citenamefont {Zhang}}]{liu2022generation}%
  \BibitemOpen
  \bibfield  {author} {\bibinfo {author} {\bibfnamefont {D.-N.}\ \bibnamefont {Liu}}, \bibinfo {author} {\bibfnamefont {J.-Y.}\ \bibnamefont {Zheng}}, \bibinfo {author} {\bibfnamefont {L.-J.}\ \bibnamefont {Yu}}, \bibinfo {author} {\bibfnamefont {X.}~\bibnamefont {Feng}}, \bibinfo {author} {\bibfnamefont {F.}~\bibnamefont {Liu}}, \bibinfo {author} {\bibfnamefont {K.-Y.}\ \bibnamefont {Cui}}, \bibinfo {author} {\bibfnamefont {Y.-D.}\ \bibnamefont {Huang}},\ and\ \bibinfo {author} {\bibfnamefont {W.}~\bibnamefont {Zhang}},\ }\bibfield  {title} {\bibinfo {title} {Generation and dynamic manipulation of frequency degenerate polarization entangled bell states by a silicon quantum photonic circuit},\ }\href@noop {} {\bibfield  {journal} {\bibinfo  {journal} {Chip}\ }\textbf {\bibinfo {volume} {1}},\ \bibinfo {pages} {100001} (\bibinfo {year} {2022}{\natexlab{b}})}\BibitemShut {NoStop}%
\bibitem [{\citenamefont {Zhang}\ \emph {et~al.}(2019)\citenamefont {Zhang}, \citenamefont {Haw}, \citenamefont {Cai}, \citenamefont {Xu}, \citenamefont {Assad}, \citenamefont {Fitzsimons}, \citenamefont {Zhou}, \citenamefont {Zhang}, \citenamefont {Yu}, \citenamefont {Wu} \emph {et~al.}}]{zhang2019integrated}%
  \BibitemOpen
  \bibfield  {author} {\bibinfo {author} {\bibfnamefont {G.}~\bibnamefont {Zhang}}, \bibinfo {author} {\bibfnamefont {J.~Y.}\ \bibnamefont {Haw}}, \bibinfo {author} {\bibfnamefont {H.}~\bibnamefont {Cai}}, \bibinfo {author} {\bibfnamefont {F.}~\bibnamefont {Xu}}, \bibinfo {author} {\bibfnamefont {S.}~\bibnamefont {Assad}}, \bibinfo {author} {\bibfnamefont {J.~F.}\ \bibnamefont {Fitzsimons}}, \bibinfo {author} {\bibfnamefont {X.}~\bibnamefont {Zhou}}, \bibinfo {author} {\bibfnamefont {Y.}~\bibnamefont {Zhang}}, \bibinfo {author} {\bibfnamefont {S.}~\bibnamefont {Yu}}, \bibinfo {author} {\bibfnamefont {J.}~\bibnamefont {Wu}}, \emph {et~al.},\ }\bibfield  {title} {\bibinfo {title} {An integrated silicon photonic chip platform for continuous-variable quantum key distribution},\ }\href@noop {} {\bibfield  {journal} {\bibinfo  {journal} {Nat. Photonics}\ }\textbf {\bibinfo {volume} {13}},\ \bibinfo {pages} {839} (\bibinfo {year} {2019})}\BibitemShut {NoStop}%
\bibitem [{\citenamefont {Llewellyn}\ \emph {et~al.}(2020)\citenamefont {Llewellyn}, \citenamefont {Ding}, \citenamefont {Faruque}, \citenamefont {Paesani}, \citenamefont {Bacco}, \citenamefont {Santagati}, \citenamefont {Qian}, \citenamefont {Li}, \citenamefont {Xiao}, \citenamefont {Huber} \emph {et~al.}}]{llewellyn2020chip}%
  \BibitemOpen
  \bibfield  {author} {\bibinfo {author} {\bibfnamefont {D.}~\bibnamefont {Llewellyn}}, \bibinfo {author} {\bibfnamefont {Y.}~\bibnamefont {Ding}}, \bibinfo {author} {\bibfnamefont {I.~I.}\ \bibnamefont {Faruque}}, \bibinfo {author} {\bibfnamefont {S.}~\bibnamefont {Paesani}}, \bibinfo {author} {\bibfnamefont {D.}~\bibnamefont {Bacco}}, \bibinfo {author} {\bibfnamefont {R.}~\bibnamefont {Santagati}}, \bibinfo {author} {\bibfnamefont {Y.-J.}\ \bibnamefont {Qian}}, \bibinfo {author} {\bibfnamefont {Y.}~\bibnamefont {Li}}, \bibinfo {author} {\bibfnamefont {Y.-F.}\ \bibnamefont {Xiao}}, \bibinfo {author} {\bibfnamefont {M.}~\bibnamefont {Huber}}, \emph {et~al.},\ }\bibfield  {title} {\bibinfo {title} {Chip-to-chip quantum teleportation and multi-photon entanglement in silicon},\ }\href@noop {} {\bibfield  {journal} {\bibinfo  {journal} {Nat. Phys.}\ }\textbf {\bibinfo {volume} {16}},\ \bibinfo {pages} {148} (\bibinfo {year} {2020})}\BibitemShut {NoStop}%
\end{thebibliography}%

\end{document}



\title{Supplementary Information: Ultracompact Low-Loss Grating Couplers}%

\author{Shiang-Yu Huang}
\email{shiang-yu.huang@fmq.uni-stuttgart.de}
\affiliation{Institute for Functional Matter and Quantum Technologies, University of Stuttgart, 70569 Stuttgart, Germany} 

\author{Jonas Zatsch}
\affiliation{Institute for Functional Matter and Quantum Technologies, University of Stuttgart, 70569 Stuttgart, Germany} 

\author{Tim Engling}
\affiliation{Institute for Functional Matter and Quantum Technologies, University of Stuttgart, 70569 Stuttgart, Germany} 

\author{Jeldrik Huster}
\affiliation{Institute for Functional Matter and Quantum Technologies, University of Stuttgart, 70569 Stuttgart, Germany} 

\author{Stefanie Barz}
\affiliation{Institute for Functional Matter and Quantum Technologies, University of Stuttgart, 70569 Stuttgart, Germany}
\affiliation{Center for Integrated Quantum Science and Technology (IQST), University of Stuttgart, 70569 Stuttgart, Germany}



\maketitle

\newpage

\section{Modification of the final design of topology-optimized 2D grating coupler}
\begin{figure*}[h!]
  \centering
  \includegraphics[scale=0.6]{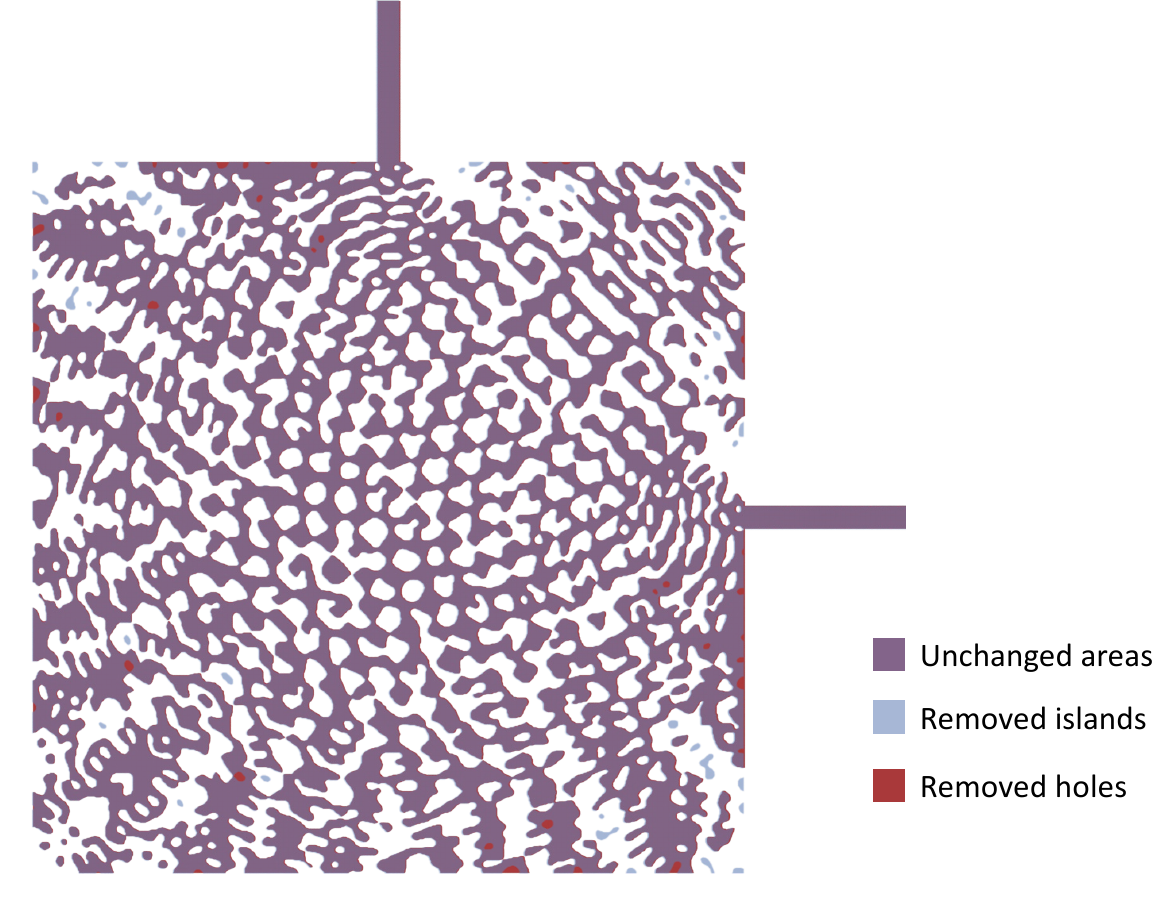}
  \caption{
  Sketch of the modification of the topology-optimized 2D grating coupler.
  }
  \label{fig:2DGC modification}
\end{figure*}
\newpage
\section{Experimental Setup for Characterization of Topology-Optimized Grating Coupler}
\begin{figure*}[h!]
  \centering
  \includegraphics[scale=0.7]{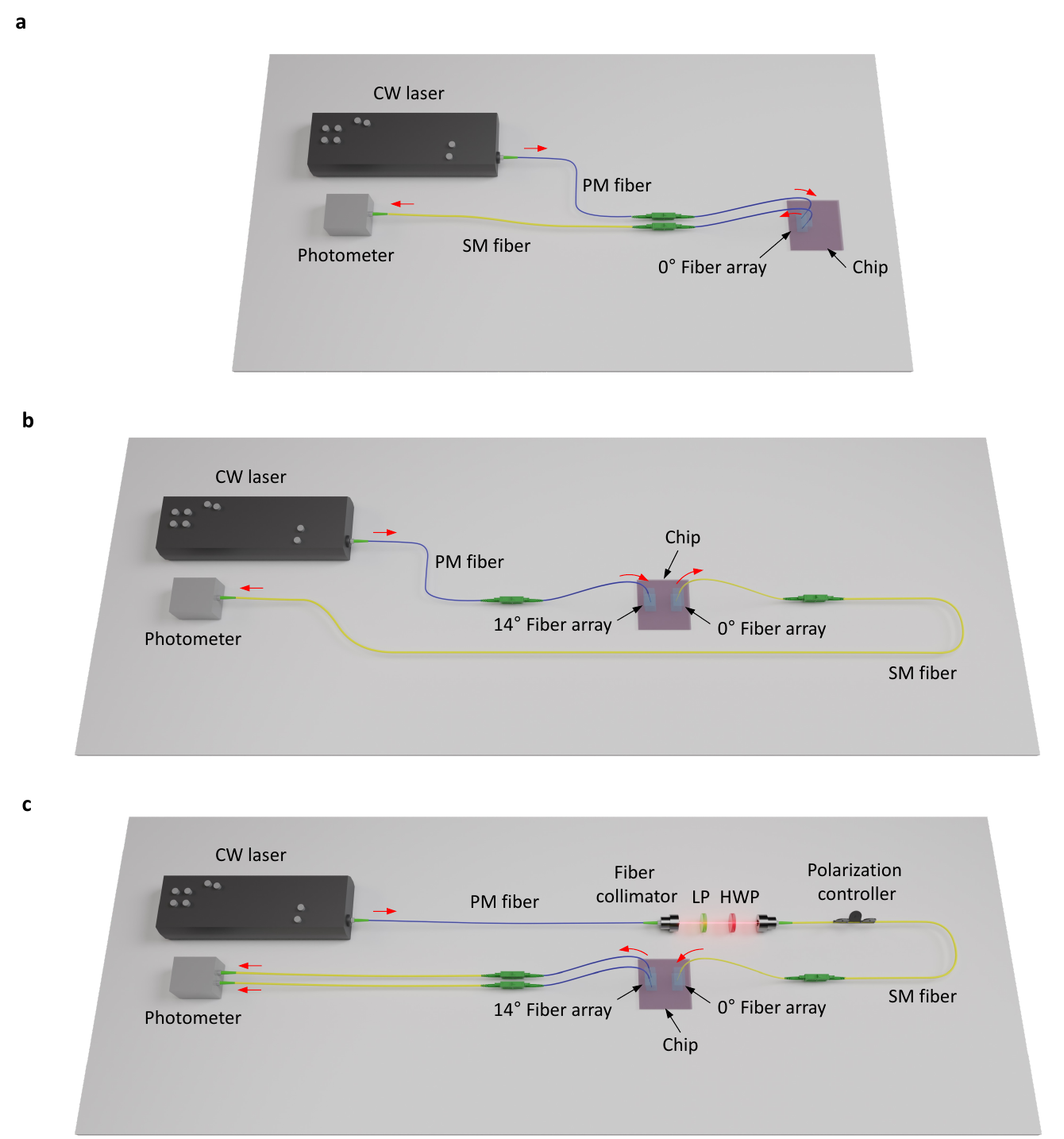}
  \caption{
  Three configurations of the experimental setup for the characterization of the topology-optimized 1D and 2D grating couplers.
  (a) The configuration for characterizing the coupling efficiency of the topology-optimized 1D grating coupler.
  (b) The configuration for characterizing the coupling efficiency of the topology-optimized 2D grating coupler.
  (c) The configuration for measuring the polarization extinction ratio and the dependency of the measured power on the input polarization of the topology-optimized 2D grating coupler.
  The red arrows indicate the propagation direction of the light.
  CW: continuous-wave.
  LP: linear polarizer.
  HWP: half-wave plate.
  SM: single mode.
  PM: polarization maintaining.
  }
  \label{fig:exp setup}
\end{figure*}

The experimental setup that uses a tunable continuous-wave (CW) laser has three different configurations tailored to the characterizations mentioned in the main text (Fig.~\ref{fig:exp setup}).
The first configuration is for characterizing the coupling efficiency of the topology-optimized 1D grating coupler (Fig.~\ref{fig:exp setup}(a)).
Light is coupled into and out of the chip via a 0\textdegree-angled polarization maintaining (PM) fiber array.

The second configuration is for characterizing the coupling efficiency of the topology-optimized 2D grating coupler (Fig.~\ref{fig:exp setup}(b)).
Light is sent to the conventional 1D grating couplers that connect to the two arms of the topology-optimized 2D grating coupler via a 14\textdegree-angled PM fiber array.
After passing through the chip, the light is coupled out from the 2D grating coupler, into a 0\textdegree-angled single-mode (SM) fiber array and is then sent to the photometer.

The third configuration is for measuring (i) the polarization extinction ratio and (ii) the dependency of the coupled power on the input polarization of the topology-optimized 2D grating coupler (Fig.~\ref{fig:exp setup}(c)).
In the first case, the linear polarizer (LP) is first adjusted to clean the polarization and maximize the power of the light going into the 0\textdegree-angled SM fiber array.
The half-wave plate (HWP) is inserted without rotation.
For each wavelength point, the light is sent into the topology-optimized 2D grating coupler and the power from both arms is maximized successively by adjusting a manual fiber polarization controller.
Finally, the power of the light from both arms is recorded by the photometer for calculating the polarization extinction ratio.
In the second case, the wavelength of the CW laser is first set to be 1550 nm. 
The angle of the LP and the configuration of the fiber polarization controller are fixed, and the power of light from both arms is recorded by the photometer when the HWP angle is swept.

\clearpage
\section{Displacement of the grating patterns in the topology-optimized grating couplers} 
\begin{figure*}[h!]
  \centering
  \includegraphics[scale=0.7]{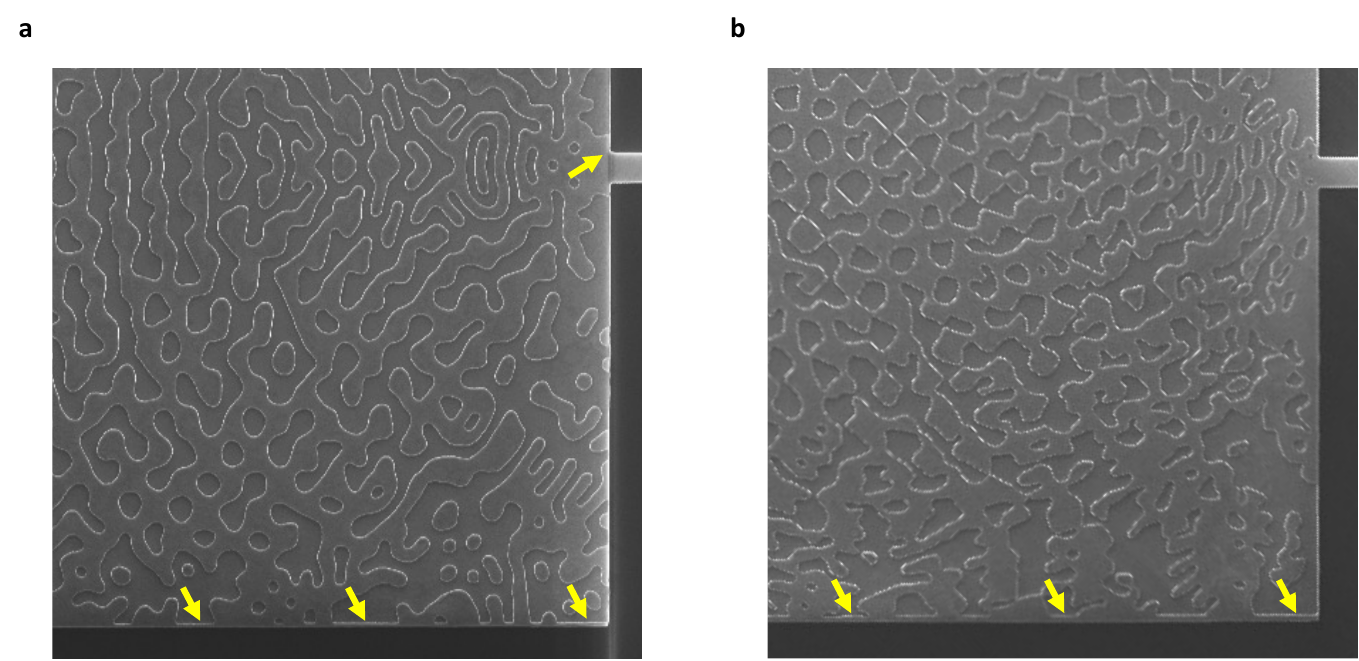}
  \caption{
  Enlarged view of the SEM images in the main text.
  (a) Topology-optimized 1D grating coupler.
  (b) Topology-optimized 2D grating coupler.
  }
  \label{fig:transmission efficiency}
\end{figure*}

\clearpage
\section{Details of the Fitting curves for the Measured power dependency on the HWP angle} 

\setlength{\tabcolsep}{12pt} 

\begin{table}[h]
    \centering
    \caption{Fitted parameters of the curves.}
    \label{tab:fit_params}
    \begin{tabular}{lcccc}
        \hline
        Curve & Amplitude (a.u.) & Period (\textdegree) & Offset (a.u.) & Initial phase (\textdegree)\\
        \hline
        X arm (orange)   & 0.4923(3) & 90.0047(24) & 0.4923(3) & $-0.4899(2)$ \\
        Y arm (blue)     & $-0.4746(6)$ & 90.0244(2) & 0.4747(6) & $-0.5157(2)$ \\
        Sum (black)      & 0.0205(8) & 92.0480(4252) & 0.9636(6) & $-0.3258(326)$ \\
        \hline
    \end{tabular}
\end{table}
